\documentclass[superscriptaddress,showpacs,showkeys,twocolumn,prb,aps,reprint,a4paper,floatfix]{revtex4-2}
\usepackage[english]{babel}
\usepackage[utf8]{inputenc}
\usepackage[colorinlistoftodos, color=green!40, prependcaption]{todonotes}
\usepackage[xetex, pdftitle={Article}, pdfauthor={Author}]{hyperref} % For hyperlinks in the PDF

\usepackage{xcolor}

\usepackage{amsmath}
\usepackage{amsfonts}

\usepackage{graphicx}
\usepackage{subcaption}
\usepackage{float}
\usepackage{caption}

\usepackage{tikz}
\usetikzlibrary{shapes.geometric, arrows, positioning}
\tikzstyle{startstop} = [rectangle, rounded corners, minimum width=2cm, minimum height=1cm, text centered, draw=black, fill=red!30]
\tikzstyle{process} = [rectangle, minimum width=2cm, minimum height=1cm, text centered, draw=black, fill=orange!30]
\tikzstyle{thermal} = [rectangle, minimum width=2cm, minimum height=1cm, text centered, draw=black, fill=blue!30]
\tikzstyle{arrow} = [thick,->,>=stealth]

\begin{document}
\graphicspath{{Figs/}}

\title{Quasiparticle Quality Factors in Superconducting Resonators: Effects of Bath Temperature and Readout Power}

\author{Zhenyuan Sun}
    \email{zs311@cantab.ac.uk}
    \affiliation{School of Integrated Circuits, Tsinghua University,  Beijing, 100084, China}
    \affiliation{Cavendish Laboratory, JJ Thomson Avenue, Cambridge, CB3 0HE, United Kingdom.}
\author{S Withington}
   \affiliation{Clarendon Laboratory, Parks Road, Oxford, OX1 3PU, United Kingdom.}
\author{Songyuan Zhao}   
   \affiliation{James Watt School of Engineering, University of Glasgow, Glasgow, UK}

\date{May 8 2026}

\begin{abstract}
The performance of superconducting resonators underpins a wide range of modern quantum technologies, yet their quality factor often deviates at low temperatures from standard Mattis-Bardeen predictions. This discrepancy is often attributed to nonthermal quasiparticles generated by microwave readout power, which limits the sensitivity of superconducting devices. We present a macroscopic model based on modified Rothwarf-Taylor equations that incorporates a power-dependent phonon generation term, providing an explicit relationship between quality factor, bath temperature and readout power. The model shows excellent agreement with temperature sweep measurements of NbN microstrip resonators with $\beta$-Ta terminations over a wide dynamic range of readout power levels, accurately capturing the transition between thermally-dominated and microwave-induced loss regimes. This framework provides a predictive tool for optimizing superconducting resonators and advancing the design of high-Q devices for quantum sensing and quantum information processing.
\end{abstract}


\maketitle

\section{Introduction}\label{sec_S0_introduction}
Superconducting resonators are essential components of a wide range of cryogenic quantum- and sensing- technologies, from sensitive detectors~\cite{zmuidzinas2012superconducting, day2006antenna, vardulakis2007superconductingastrophysics, golwala2008wimp} and parametric amplifiers~\cite{zhao2023nonlinear, zhao2024superconducting} to the control and readout of qubits~\cite{wallraff2004strong, tuokkola2025methods} and quantum memories~\cite{Sullivan_2022,Matanin_2023}. Their utility critically depends on achieving high quality factors (Q-factors) and minimal dissipation, ensuring low signal loss, high fidelity, and stable device operation.  Recent reviews emphasize how the interplay of dielectric two-level systems (TLS), non-equilibrium quasiparticles, trapped flux vortices and other residual loss mechanisms defines the practical limits of resonator performance~\cite{mcrae2020materials, gurevich2023tuning}. 

According to Mattis-Bardeen theory, the real part of the complex conductivity vanishes exponentially as the bath temperature decreases to zero~\cite{mattis1958theory}, implying that, in principle, the Q-factor should increase without bound. In practice, high-Q resonators saturate, i.e. Q-factor flattens or even reduces at low temperatures, deviating markedly from the predicted theoretical behaviour~\cite{zmuidzinas2012superconducting, gao2008physics, mauskopf2018transition}. This discrepancy is often attributed to non-equilibrium quasiparticle populations generated by the readout microwave power, a phenomenon known as quasiparticle heating (QPH)~\cite{thomas2020nonlinear},  which injects a finite density of excitations even at the lowest temperature. In addition, losses from TLS can further suppress the Q-factor as temperature decreases~\cite{skyrme2023superconducting}. 

Microscopic models have been previously developed to describe the interactions between quasiparticles and phonons, along with their energy spectra in the presence of sub-gap microwave photons~\cite{chang1977kinetic, goldie2012non}. In terms of resonator dynamics, the resulting behaviour is captured by a reduced model in which quasiparticles are characterised by an effective temperature exceeding the physical bath temperature. In this framework, the power dissipated by readout microwave signal is regarded as effectively heating the quasiparticles~\cite{de2010readout}, and a steady state is established when the heating power is balanced by cooling through energy transfer to phonons~\cite{goldie2012non, guruswamy2015nonequilibrium}. Such an electrothermal model has successfully described both large- and small-signal device behaviour~\cite{thompson2013dynamical, thomas2015electrothermal, guruswamy2017electrothermal}.

In this work, we present a macroscopic model based on the Rothwarf-Taylor equations~\cite{rothwarf1967measurement}, a well-established phenomenological framework for describing the coupled dynamics of non-equilibrium quasiparticles and high-energy phonons in superconductors. We extend it by introducing an additional quasiparticle generation term that depends explicitly on the microwave readout power. This approach provides a robust and tractable method for solving the quasiparticle population kinetics by directly coupling generation and recombination processes.  Compared to earlier effective-temperature models~\cite{goldie2012non, Antonenko_2026}, our formulation explicitly tracks the total quasiparticle density rather than invoking an effective temperature. This offers a direct connection to the underlying microscopic scattering rates and facilitates a clearer interpretation of power- and temperature-dependent losses. The resulting framework provides both analytical and numerical means for calculating the quality factor as a function of bath temperature and readout power, enabling direct comparison with experimental data.  

Loss mechanisms in superconducting resonators are dependent on the operating regime: TLS loss typically dominates in the low readout power regime, whereas QPH becomes increasingly significant at higher readout powers. The model presented here primarily addresses the temperature- and power-dependent quasiparticle dissipation that governs performance after TLS-related losses have saturated. 

In addition to the theoretical work, we fabricated NbN microstrip resonators and performed systematic bath temperature and readout power sweeps to evaluate the model. The measurements spanned a readout power dynamic range of up to 40\,dB and temperatures reaching approximately $T_c/2$. Across this experimental parameter space, the extracted quality factors showed strong agreement with predictions of our modified Rothwarf-Taylor framework, capturing the low-temperature plateau, crossover regime, and high-temperature decay. Overall, this modified Rothwarf-Taylor quasiparticle framework provides a predictive tool for guiding the design and operation of high-Q superconducting resonators for quantum sensing and quantum information processing.

\section{Theoretical model}\label{sec_S2_theoretical_model}
Our approach builds on the Rothwarf-Taylor equations~\cite{rothwarf1967measurement} by introducing a quasiparticle generation-rate term $\Gamma_r$, at which pair-breaking phonons are generated from readout power $P_r$,  in Eq.~\ref{eqn:S1_rothwarf_taylor_equation_2}
\begin{equation}\label{eqn:S1_rothwarf_taylor_equation_1}
\frac{\partial n_{qp}}{\partial t} = \frac{2}{\tau_{pb}}n_\omega - Rn^2_{qp},
\end{equation}
\begin{equation}\label{eqn:S1_rothwarf_taylor_equation_2}
\frac{\partial n_{\omega}}{\partial t} = -\frac{1}{\tau_{pb}}n_\omega + \frac{R}{2} n^2_{qp} - \frac{1}{\tau_{l}} [n_\omega - n_{\omega, th}] + \Gamma_r.
\end{equation}
$n_{qp}$ is the quasiparticle number density, $n_{\omega}$ is the number density of phonons with energy greater than the pair-breaking threshold in the same active volume $V$ of the resonator and $n_{\omega,th}$ is the value of $n_{\omega}$ in thermal equilibrium with no forcing ($\Gamma_r = 0$).  
%$\Gamma_r$ is the rate at which pair-breaking phonons are generated from readout power $P_r$, 
$\tau_{pb}$ is the pair-breaking time, $R$ is the quasiparticle recombination rate and $\tau_l$ is the lifetime of a pair-breaking phonon in the absence of interactions with the quasiparticle system. 

In our model, we assume steady-state operation ($\partial n_{qp} / \partial t = 0$, $\partial n_{\omega} / \partial t = 0$) and negligible direct pair-breaking via microwave readout photons since the readout frequency is usually well below the pair-breaking frequency threshold. The effect of readout signal enters through $\Gamma_r$ in Eq.~\ref{eqn:S1_rothwarf_taylor_equation_2}, representing the conversion of dissipated microwave power into pair-breaking phonons via quasiparticle-phonon elastic scattering~\cite{guruswamy2015nonequilibrium}.  Eliminating $n_\omega$ by substituting Eq.~\ref{eqn:S1_rothwarf_taylor_equation_2} into Eq.~\ref{eqn:S1_rothwarf_taylor_equation_1} yields 
\begin{equation}
R n_{qp}^2 = \frac{2 \tau_l}{\tau_{pb}} \left[\Gamma_r + \frac{1}{\tau_l}n_{\omega, th}  \right]\,,
\end{equation}
which is then simplified as
\begin{equation}\label{eqn:S1_nqp_steady_state_vs_n_omega_th_with_forcing_simplification}
R [n_{qp}^2 -  n_{qp,th}^2]= \frac{2\tau_l}{\tau_{pb}} \Gamma_r\,.
\end{equation}
In the low readout-power limit $\Gamma_r = 0$, the quasiparticle density $n_{qp}$ takes on the thermal equilibrium value $n_{qp, th}$.  Eq.~\ref{eqn:S1_nqp_steady_state_vs_n_omega_th_with_forcing_simplification} shows $n_{qp}$ is determined by balancing the recombination rate with the generation rate due to the readout signal. It also demonstrates that when $\tau_l \ll \tau_{pb}$, phonons have decayed before they can generate quasiparticles and thus the effective rate at which quasiparticles are generated by phonons is reduced compared to the intrinsic limit. 

We assume $\Gamma_r = \eta P_{qp} / V \Delta$ (the energy gap $\Delta \approx 1.76 k_B T_c$) is proportional to the power $P_{qp}$ dissipated into quasiparticle system by readout signal, and $\eta$ is generation efficiency.  The internal losses in the resonator are split into those associated with quasiparticles and all other losses, i.e. $Q_i^{-1} = Q_{qp}^{-1} + Q_{other}^{-1}$. Accordingly, $P_{qp} = P_{diss}{Q_i}/{Q_{qp}} $, where $P_{diss}$ is the total power dissipated in the resonator, given by 
\begin{equation}
P_{diss}  =  \frac{2Q_r}{Q_c} \frac{1}{1 + (2Q_r x)^2} \frac{Q_r}{Q_i} P_r\,.
\end{equation}

The total quality factor $Q_r$ is given by $Q_r^{-1} =Q_i^{-1} + Q_c^{-1}$, where $Q_c$ is the coupling quality factor. $x$ is the fractional frequency detuning given by $x = (f - f_r)/f_r$, where $f$ is the frequency of the readout signal and $f_r$ is the resonance frequency. The Mattis-Bardeen theory predicts $Q_{qp}$ to be inversely proportional to $n_{qp}$~\cite{mccarrick2014horn}. This dependence allows the introduction of a scaling parameter $n_\ast$ which encapsulates the effects of temperature, frequency, and resonator geometry, leading to the convenient parametrization $Q_{qp} = Q_c{n_\ast}/{n_{qp}} $. We assume $x = 0$ subsequently and $\Gamma_r = \eta P_{qp} / V \Delta$ becomes

\begin{equation}\label{eqn:S1_Gamma_r_vs_Pr}
\Gamma_r = \frac{n_{qp} n_\ast}{[n_\ast (1 + Q_c/Q_{other}) + n_{qp}]^2} \frac{2 \eta P_r}{V \Delta}.
\end{equation}

Substituting Eq.~\ref{eqn:S1_Gamma_r_vs_Pr} into Eq.~\ref{eqn:S1_nqp_steady_state_vs_n_omega_th_with_forcing_simplification} yields the governing equation
\begin{equation}\label{eqn:S1_normalised_governing_equation}
\resizebox{0.95\width}{!}{$\begin{aligned}
u^4 &+ 2 a u^3 + (a^2 - u_{th}^2) u^2 - (2 a u_{th}^2 + \gamma) u - a^2 u_{th}^2 = 0\\
&u = \frac{n_{qp}}{n_{\ast}}, \,\,\,\,\,\,\,\, u_{th} = \frac{n_{qp,th}}{n_{\ast}}, \,\,\,\,\,\,\,\, a = 1 + \frac{Q_c}{Q_{other}},
\end{aligned}$}
\end{equation}
where 
\begin{equation}\label{eqn:norm_Power}
	\gamma = 4 \eta \tau_l P_r / \tau_{pb} R V \Delta n_{\ast}^2
\end{equation} is the normalised applied power.  

For $T_b \ll T_c$, Mattis-Bardeen theory predicts the number density of thermally excited quasiparticles to be~\cite{mattis1958theory, gao2008equivalence}
\begin{equation}\label{eqn:S1:nqpth_vs_bath_temperature}
n_{qp,th} (T_b) = 2 n_0 \sqrt{2 \pi k_B T_b \Delta} e^{- \frac{\Delta}{k_B T_b}},
\end{equation}
where $n_0$ is the single spin density of states at Fermi surface of the superconductor,  and $k_B$ is the Boltzmann constant. Comparing $Q_{qp} = \frac{n_\ast}{n_{qp}} Q_c$ with the exact result given in Refs.\cite{mccarrick2014horn, zmuidzinas2012superconducting}, i.e. $Q_{qp} = \frac{4 n_0 \Delta}{2 \alpha S_1(f) n_{qp}}$, yields 
\begin{equation}\label{eqn:S1:nast_specific_formation}
n_\ast = \frac{\pi n_0 \sqrt{2 \pi k_B T_b \Delta}}{2 \alpha Q_c \sinh(\xi) K_0(\xi)},
\end{equation}
where $S_1(f) = \frac{2}{\pi} \sqrt{\frac{2 \Delta}{\pi k_B T_b}} \sinh\left( \xi \right) K_0\left( \xi \right)$, $\xi = \frac{hf}{2k_B T_b}$, $f$ is the readout frequency, $\alpha$ is kinetic inductance fraction and $K_0$ is the modified Bessel function of the second kind. From a device-design perspective, it is useful to note that $n_\ast \propto \sqrt{T_c}$ (through $\sqrt{\Delta}$ dependence) and $1/Q_c$. Normalising $n_{qp,th}$ by $n_{\ast}$ in Eq.~\ref{eqn:S1:nqpth_vs_bath_temperature} and substituting $n_\ast$ by Eq.~\ref{eqn:S1:nast_specific_formation} yield
\begin{equation}\label{eqn:S1:uth_VS_Tc}
u_{th} =  \frac{4 \alpha Q_c }{\pi} e^{-1.76 \frac{T_c}{T_b}}\sinh \left( \frac{h f}{2 k_B T_b} \right) K_0 \left( \frac{h f}{2 k_B T_b} \right) .
\end{equation}
Eq.~\ref{eqn:S1:uth_VS_Tc} shows that each bath temperature corresponds to a thermal quasiparticle density $u_{th}$. For any given combination of $T_b$ and $P_r$, Eq.~\ref{eqn:S1_normalised_governing_equation} can be solved to determine the normalised quasiparticle density $u$, thereby establishing the explicit dependence of the normalised quasiparticle density $u$ on the bath temperature $T_b$ and the readout power $P_r$.

In experimental characterisation of resonator dynamics, it is convenient to eliminate $n_{\ast}$ by substituting Eq.~\ref{eqn:S1:nast_specific_formation} into Eq.~\ref{eqn:norm_Power}. This groups the material and device parameters $n_0$, $\tau_{pb}$, $\tau_l$, $R$, $V$, and $\eta$ into a single power scale $P_c$. The resulting expression is
\begin{equation}\label{eqn:S1:gamma}
\gamma = (\alpha Q_c)^2 \frac{P_{r'}}{P_c}\frac{T_c}{T_b} \sinh^2 \left( \frac{h f}{2 k_B T_b} \right) K_0^2 \left( \frac{h f}{2 k_B T_b} \right)\,,
\end{equation} 
where $P_{r'} = L_0 P_r$ denotes the power to a convenient measurement plane,  $P_r$ denotes the power at the resonator,  $L_0$ is the linear path loss factor between the measurement plane and the resonator device and the corresponding power scale $P_c$ is given by
\begin{equation}
P_c = \frac{\pi^3 k_B n_0^2 \tau_{pb} L_0 R V \Delta^2 T_c}{8 \eta  \tau_l}.
\end{equation} 

In summary, the quality factor is computed numerically for each combination of bath temperature $T_b$ and readout power $P_r$ as follows. First, the normalised thermal quasiparticle density $u_{th}$ is evaluated from Eq.~\ref{eqn:S1:uth_VS_Tc}, and the normalised applied power $\gamma$ is evaluated from Eq.~\ref{eqn:S1:gamma}. These two quantities are then substituted into the quartic equation (Eq.~\ref{eqn:S1_normalised_governing_equation}), which is solved numerically for the normalised quasiparticle density $u = n_{qp}/n_*$. From the resultant $u$, $Q_{qp}$ is obtained using $Q_{qp} = Q_c/u$, and $Q_i$ is then obtained using $Q_i^{-1} = Q_{qp}^{-1} + Q_{other}^{-1}$. This procedure is repeated across the relevant parameter space of $T_b$ and $P_r$ to generate the model predictions.

\begin{figure}[htbp!]
\centering
  	\begin{tikzpicture}
  	%\vspace{-2cm}
    	\node[inner sep=0, xshift=0cm, yshift=0cm] (image) at (0,0) {\includegraphics[width=.9\linewidth]{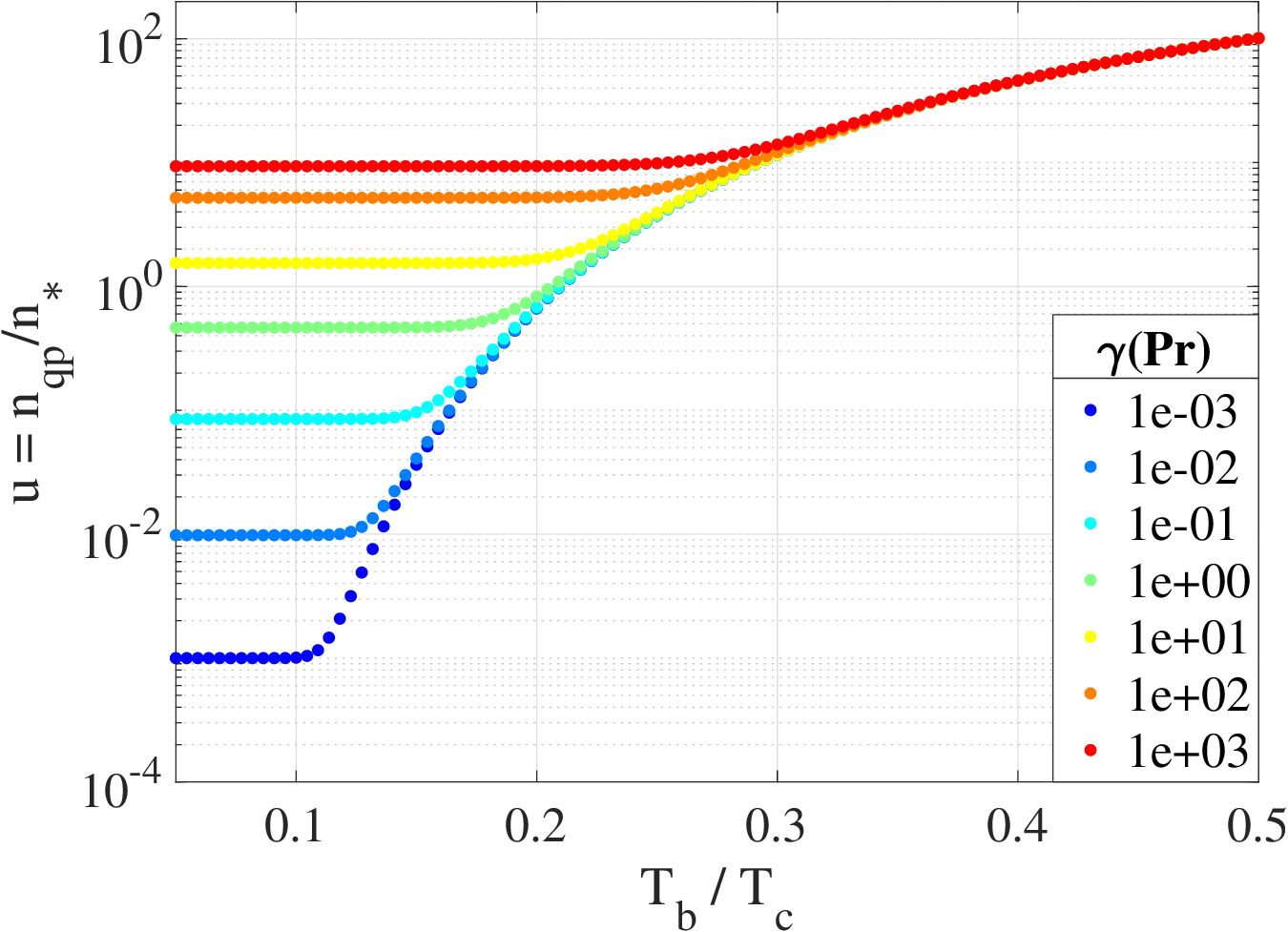}};
    	 \node[inner sep=0, xshift=0cm, yshift=-6.26cm] (image) at (0,0) {\includegraphics[width= .9\linewidth]{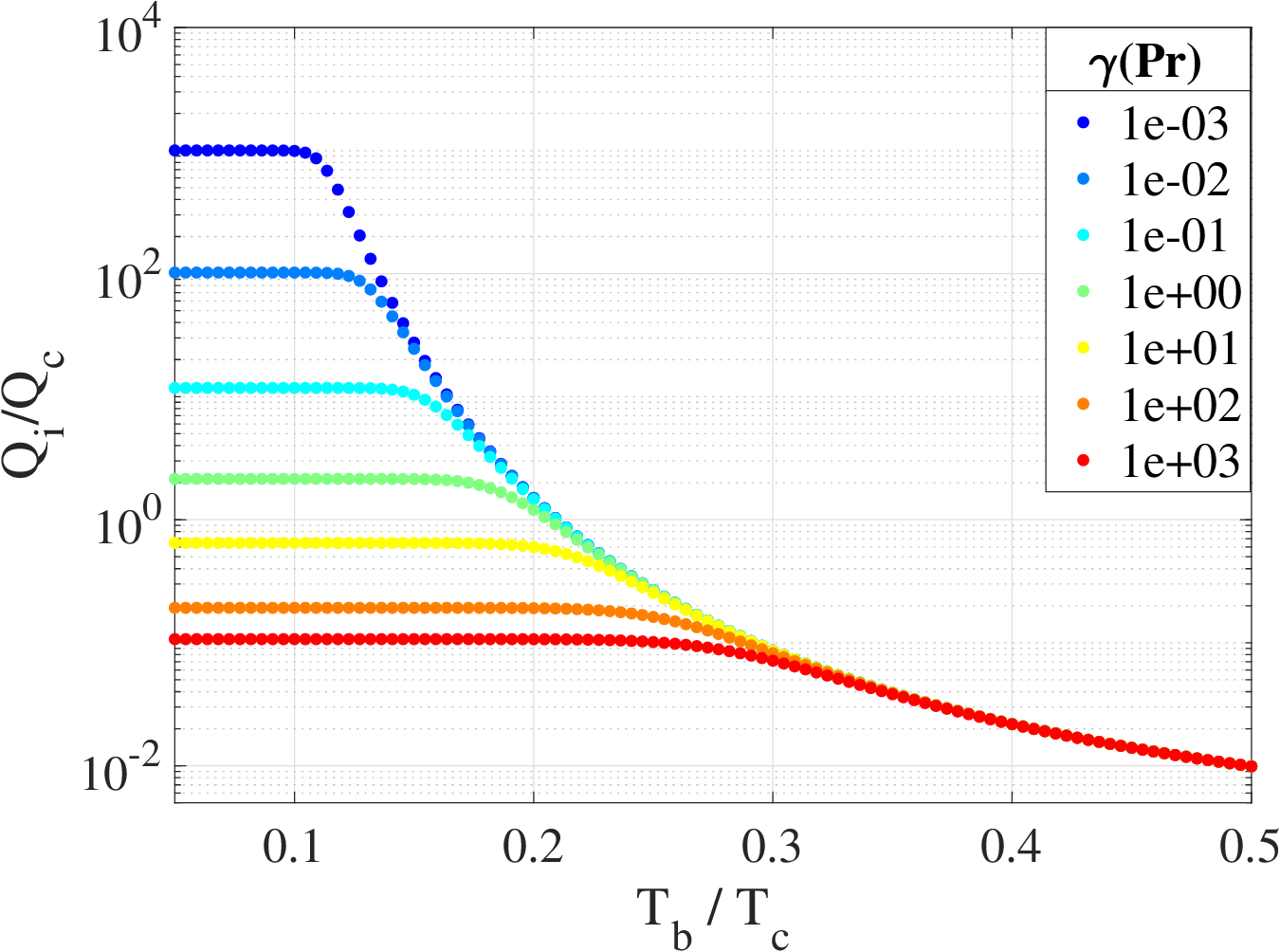}};

    	\node[text=black, scale=1, align=left] at (0.25,-3.25) {(a)};
    	\node[text=black, scale=1, align=left] at (0.25,-9.5) {(b)};
    	
 	 \end{tikzpicture}
\caption[]{\label{fig:S1_u_Qi_vs_bath_temperature_different_Pr_simulation} Dependence of $u$ and $Q_i/Q_c$ on reduced bath temperature $T_b/T_c$ at varying readout power levels. }
\end{figure}

Simulations were performed based on the framework described above, using parameters: $\alpha =$ 0.7, $f_r$ = 3.5\,GHz, and $T_c$ = 0.7\,K, which were close to those of the fabricated and measured NbN microstrip resonators discussed in Sec.\ref{Sec3_bath_temperature_sweep_measurement}.
%representative of the fabricated and measured NbN microstrip resonators discussed in Sec.\ref{Sec3_bath_temperature_sweep_measurement}:
Fig.~\ref{fig:S1_u_Qi_vs_bath_temperature_different_Pr_simulation}(a) shows the normalised quasiparticle density $u$ as a function of reduced bath temperature $T_b/T_c$ at different readout power levels. As shown in the figure, at low reduced temperatures ($T_b/T_c < 0.1$), the quasiparticle density exhibits a near-constant saturation plateau before rising with increasing temperature. The level of this residual plateau is set by the applied readout power, with higher power producing a higher baseline quasiparticle density. As the bath temperature increases further, the system enters a crossover regime in which temperature- and power-induced quasiparticle populations are comparable. At higher temperatures, thermally generated quasiparticles dominate the total density, which increases monotonically with bath temperature.

Correspondingly, Fig.~\ref{fig:S1_u_Qi_vs_bath_temperature_different_Pr_simulation}(b) shows the normalised internal quality factor $Q_i/Q_c$ as a function of reduced bath temperature $T_b/T_c$, which is directly accessible in experiment. Here we have assumed that $Q_i$ is limited by quasiparticle processes, i.e. $Q_i = Q_{qp}$. At low reduced temperatures, $Q_i/Q_c$ exhibits a high plateau whose level depends on the applied readout power, reflecting the quasiparticle population set by microwave-induced generation. As the temperature increases, the system enters a crossover regime where thermally and power-induced quasiparticles contribute comparably, followed by a monotonic decay of $Q_i/Q_c$ once thermally generated quasiparticles dominate. The behavior of the quality factor therefore provides a direct experimental probe of both the model and the underlying quasiparticle dynamics.

\section{Experimental measurements}\label{Sec3_bath_temperature_sweep_measurement}

\begin{figure}[htbp!]
\centering
  	\begin{tikzpicture}
  	%\vspace{-2cm}
    	\node[inner sep=0, xshift=0cm, yshift=0cm] (image) at (0,0) {\includegraphics[width=\linewidth]{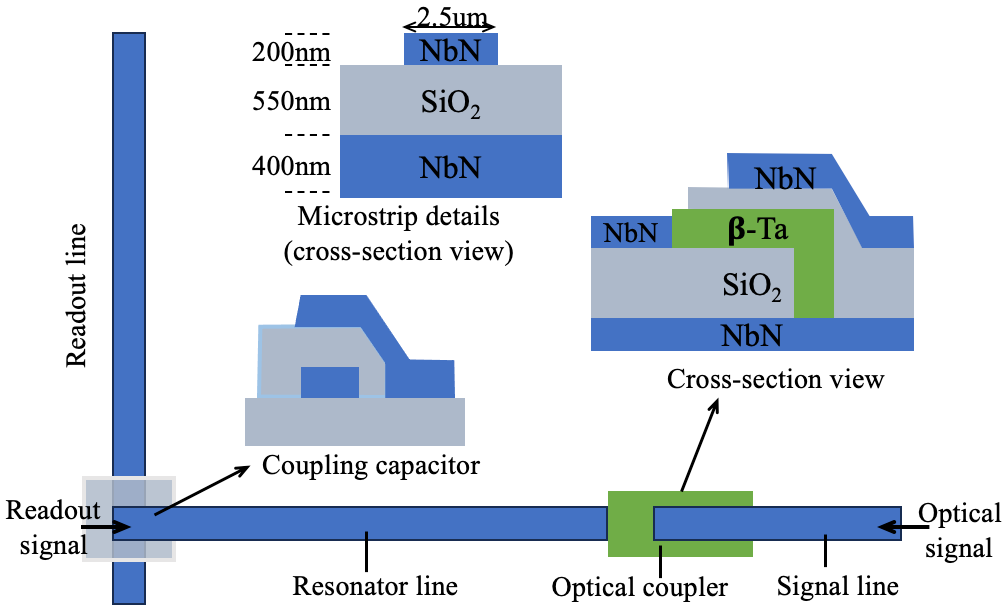}};
 	 \end{tikzpicture}
\caption[]{\label{fig:S2_microstrip_resonator} Schematic diagram of NbN resonator microstrip.}
\end{figure}

Superconducting NbN thin films were deposited and patterned to realize the microstrip resonators investigated in this study. The NbN films were grown by reactive DC magnetron sputtering and subsequently defined by reactive ion etching~\cite{glowacka2014development}. The resonators were originally developed for the CAMbridge Emission Line Surveyor (CAMELS) project~\cite{glowacka2014development, thomas2014cambridge, thomas2015progress,zengoptical} and were measured in this study to evaluate the modified Rothwarf-Taylor model presented in the previous section. Fig.~\ref{fig:S2_microstrip_resonator} shows a schematic diagram of the quarter-wavelength microstrip resonator, shorted at one end by a $\beta$-Ta via and lightly capacitively coupled to the readout line at the other. The feedline signal conductor overlaps the resonator terminus such that the $\beta$-Ta ($T_c \approx 500-800\,\mathrm{mK}$~\cite{dhundhwal2025high}) functions as the sensing material and forms the ground plane of the signal line. The NbN ($T_c \approx 11\,\mathrm{K}$) top and bottom conductors, with thicknesses of 200\,nm and 400\,nm respectively, are separated by a 550\,nm SiO$_2$ dielectric layer, and the signal-line width is 2.5\,$\mu$m. The resonator lengths were designed to lie in the range of 3.5 to 4.5\,mm. A 200\,nm SiO$_2$ layer, patterned by lift-off, was used to form the coupling capacitor. The resulting overlap between the resonator and the readout line created a parallel-plate capacitor that determined the external coupling quality factor.

The resonators were wire-bonded using aluminium wires to a gold-plated copper enclosure with SMA connector and cooled to $\sim$100\,mK in an adiabatic demagnetisation refrigerator (ADR). A high electron mobility transistor (HEMT) amplifier, mounted at the $\sim$4\,K stage, was used to amplify the output signal from the resonator. Using a vector network analyzer, frequency sweeps were performed over a range of readout power levels and bath temperatures to measure the resonator transmission response. The resulting spectra were fitted to extract the quality factors. 

\begin{table}[htbp!]
\centering
\begin{tabular}{|c|c|c|c|c|c|c|}
\hline
$P_r$ / [dBm] & -35 & -40 & -50 & -60 & -70 & -80 \\
\hline
$Q_{other} / Q_c$ (R2) &0.0545 &0.0797  &0.1812  &0.1451  &0.0792 &0.0471 \\
\hline
$Q_{other} / Q_c$ (R1) &-             &0.0606 &0.1502  &0.1592 &0.1578 &0.1116\\
\hline
\end{tabular}
\caption{\label{table_fitting_parameter_R2_R1} Best fit values of $Q_{other} / Q_c$ for bath temperature sweep data at different readout powers for R1 and R2. Additional inputs to the model include the power scaling parameter $P_c$ and the superconducting transition temperature $T_c$, both obtained from a single global fit for each device and held fixed across all bath temperatures and readout powers.  For R1 and R2, $P_c$ was determined to be $37\,\mathrm{mW}$ and $95\,\mathrm{mW}$, and $T_c$ was determined to be $0.64\,\mathrm{K}$ and $0.63\,\mathrm{K}$, respectively. The kinetic inductance fraction $\alpha$ was determined to be $0.7$ for both resonators, following the calculation detailed in Ref.~\cite{mazin2010thin}.}
\end{table}

Two resonators, labeled R1 and R2, were measured. At a bath temperature of $\sim100\,\mathrm{mK}$, they exhibited resonance frequencies of $3.7895\,\mathrm{GHz}$ and $3.4290\,\mathrm{GHz}$, and coupling quality factors $Q_c$ of $19000$ and $21000$, respectively. Further information on device design considerations can be found in reports from the CAMELS project~\cite{glowacka2014development, thomas2014cambridge, thomas2015progress,zengoptical} and additional details on the measurement system can be found in Ref.~\cite{sun2025superconducting}. The parameters used in the fitting procedure are summarised in Table~\ref{table_fitting_parameter_R2_R1} and its caption. The global parameters $P_c$ and $T_c$ were obtained from a single global fit for each device and held fixed across all bath temperatures and readout powers. For R1 and R2, $P_c$ was determined to be $37\,\mathrm{mW}$ and $95\,\mathrm{mW}$, and $T_c$ was determined to be $0.64\,\mathrm{K}$ and $0.63\,\mathrm{K}$, respectively. The kinetic inductance fraction $\alpha$ was determined to be $0.7$ for both resonators, following the calculation detailed in Ref.~\cite{mazin2010thin}.
Within this framework, $Q_{other}$ was the only free parameter allowed to vary across readout power levels, and it was determined by best fit to the measured data at each power. It represents the contribution of non-quasiparticle loss mechanisms to the measured resonator response. 

\begin{figure}[htbp!]
\centering
  	\begin{tikzpicture}
  	%\vspace{-2cm}
    	 \node[inner sep=0, xshift=0cm, yshift=-6.26cm] (image) at (0,0) {\includegraphics[width= .9\linewidth]{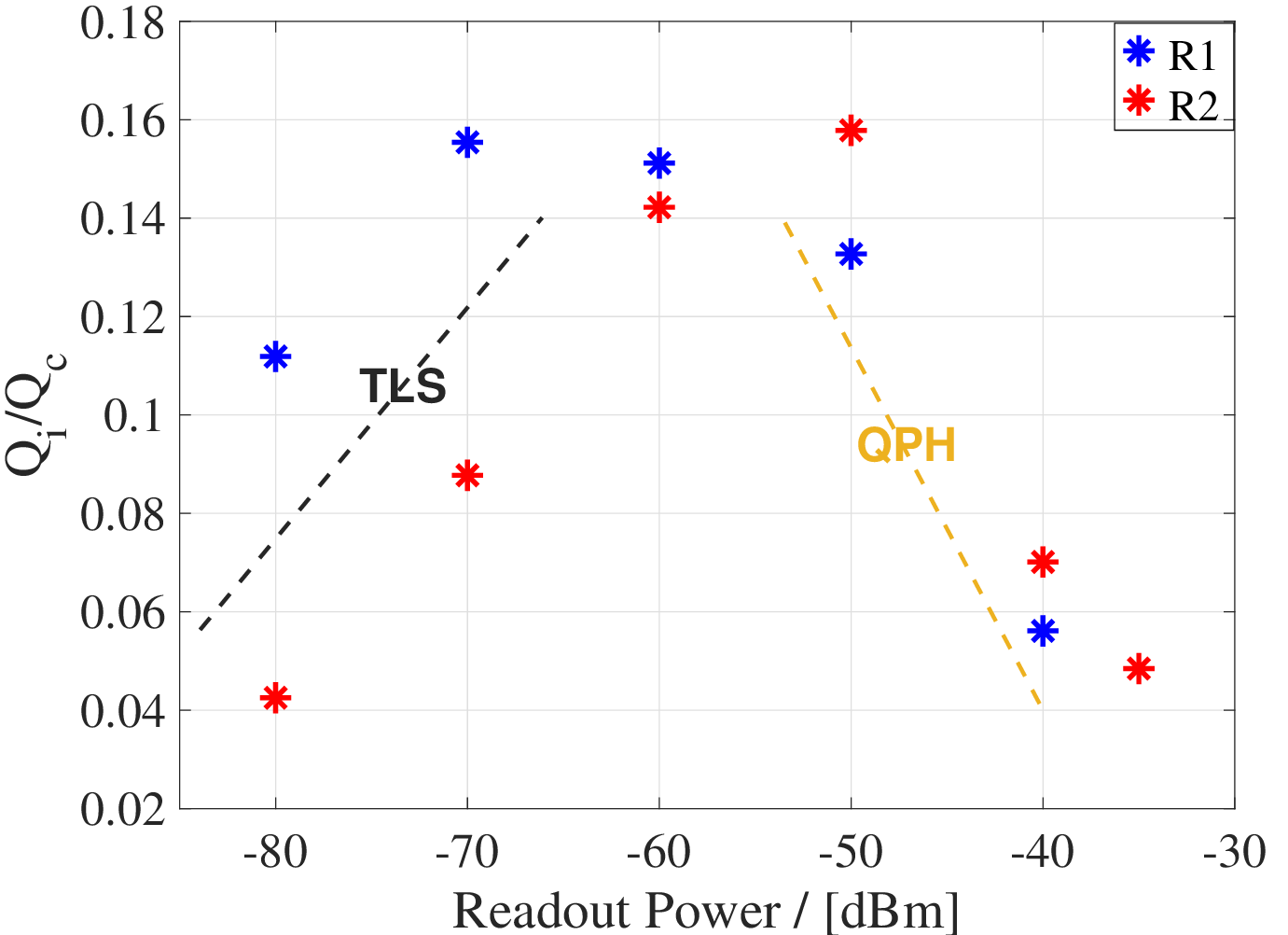}};

 	 \end{tikzpicture}
\caption[]{\label{fig:S1_measured_S21_vs_Pr_at_100mK}  Measured quality factor versus readout power at 110\,mK for NbN resonators $R_1$ and $R_2$ with $\beta$-Ta.}
\end{figure}
\begin{figure}[htbp!]
\centering
  	\begin{tikzpicture}
  	%\vspace{-2cm}
    	\node[inner sep=0, xshift=0cm, yshift=0cm] (image) at (0,0) {\includegraphics[width=.9\linewidth]{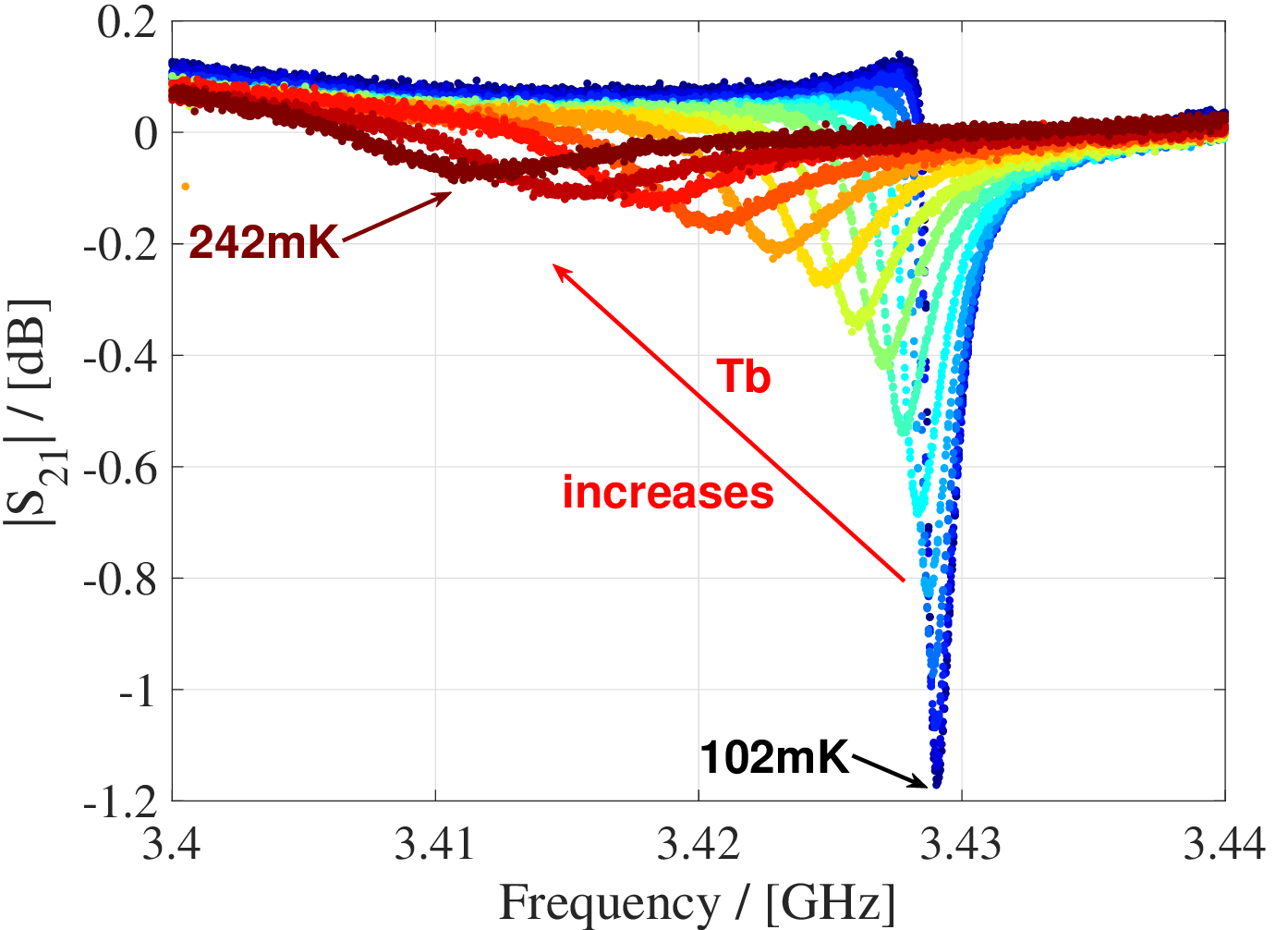}};
    	 \node[inner sep=0, xshift=0cm, yshift=-6.26cm] (image) at (0,0) {\includegraphics[width= .9\linewidth]{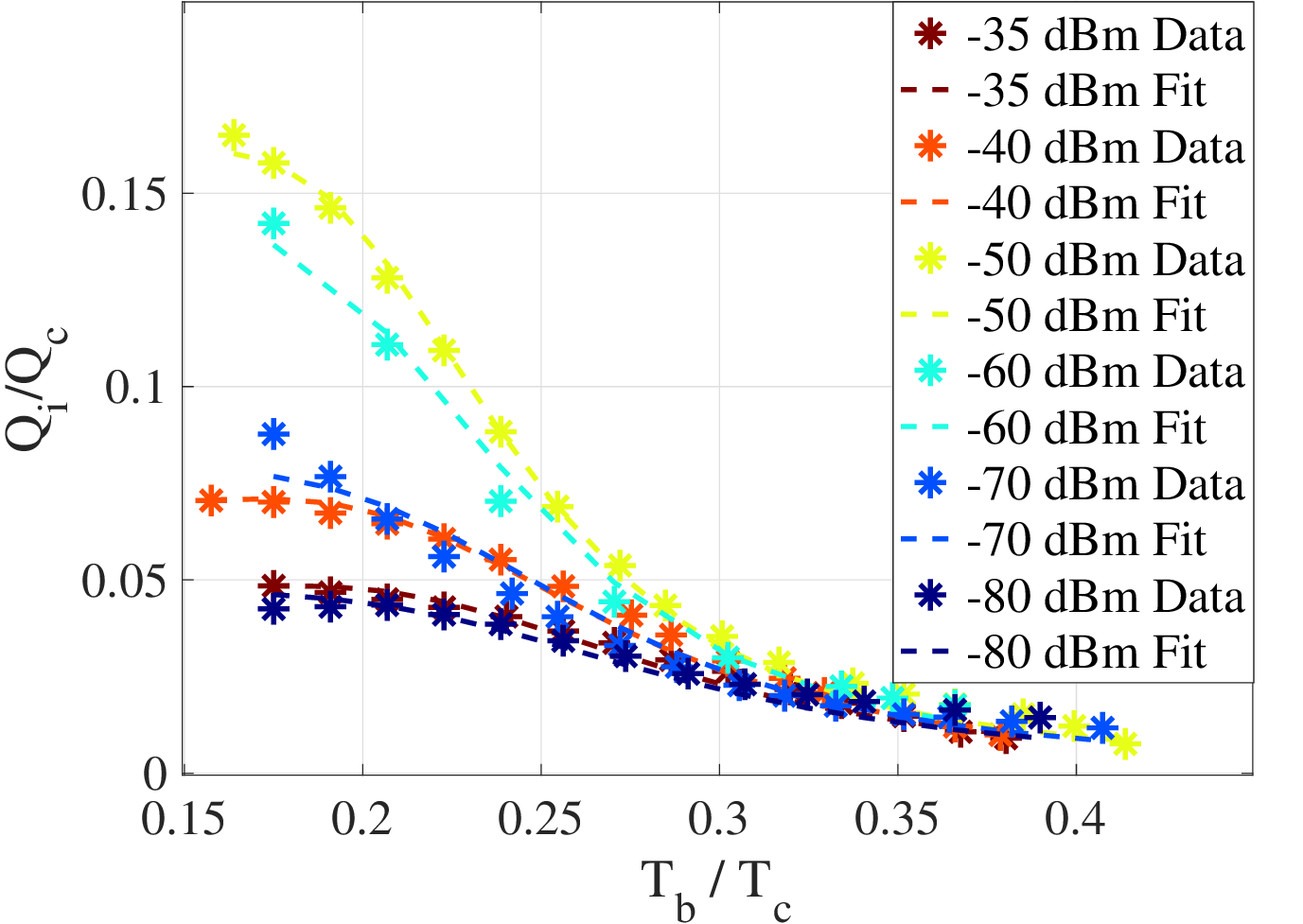}};

    \node[text=black, scale=1, align=left] at (0.25,-3.25) {(a)};
    \node[text=black, scale=1, align=left] at (0.25,-9.5) {(b)};
    	
 	 \end{tikzpicture}
\caption[]{\label{fig:S1_measured_S21_vs_bath_temperature_at_different_readout_power_R2} (a) Measured $|S_{21}|$ versus readout frequency at different bath temperatures with -50\,dBm readout power for NbN resonator $R_2$. (b) Extracted $Q$ factor as a function of bath temperature at different readout powers. The experimentally extracted $Q$ factors are shown as star markers, while the fits to the proposed model are represented by dashed lines.}
\end{figure}
\begin{figure}[htbp!]
\centering
  	\begin{tikzpicture}
    	 \node[inner sep=0, xshift=0cm, yshift=-6.26cm] (image) at (0,0) {\includegraphics[width= .9\linewidth]{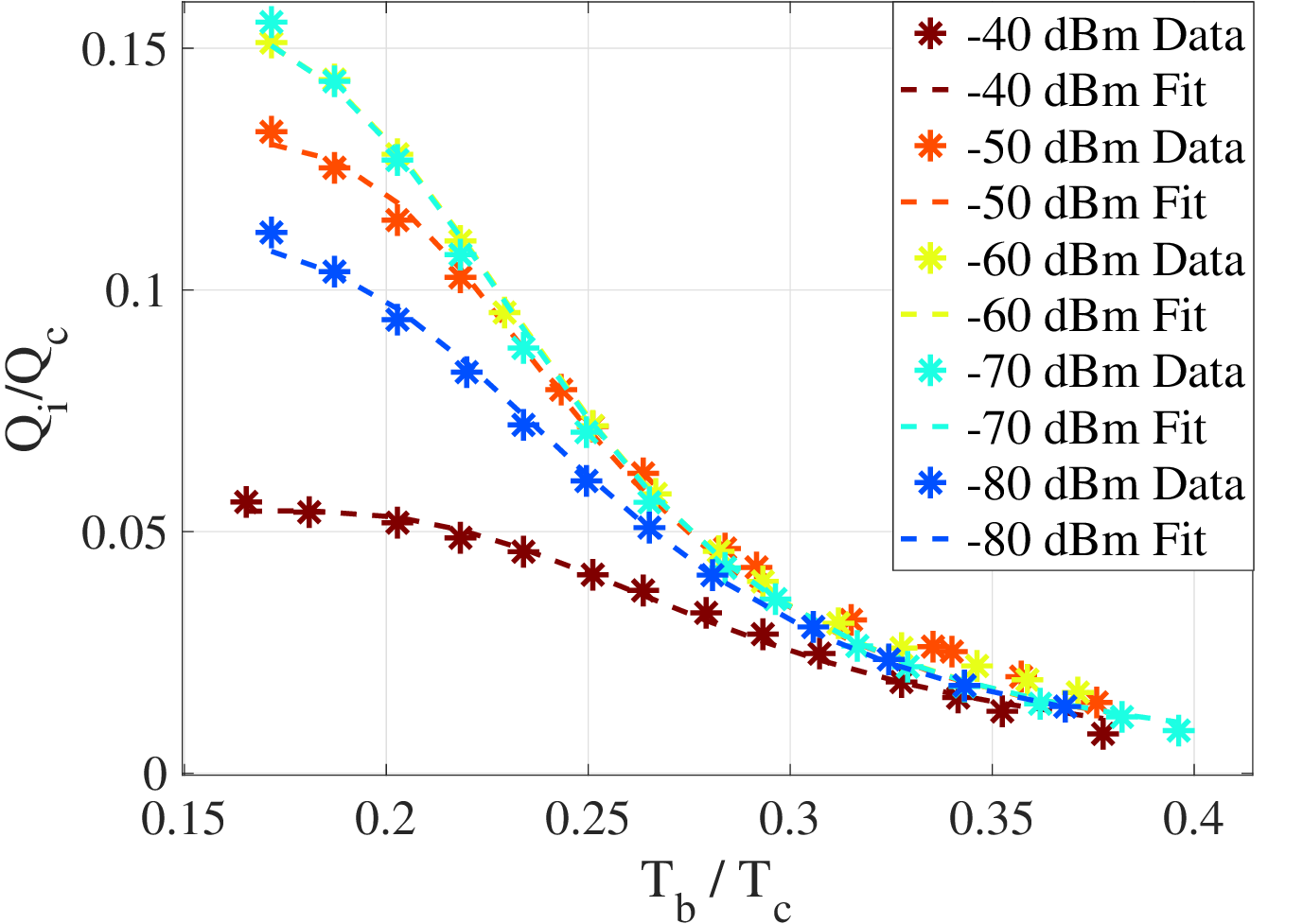}};    	
 	 \end{tikzpicture}
\caption[]{\label{fig:S1_measured_S21_vs_bath_temperature_at_different_readout_power_R1} Extracted $Q$ factors as a function of bath temperature at different readout powers for R1. The experimentally extracted $Q$ factors are shown as star markers, while the fits to the proposed model are represented by dashed lines.}
\end{figure}

Figure~\ref{fig:S1_measured_S21_vs_Pr_at_100mK} shows the normalised internal quality factor $Q_i/Q_c$ as a function of readout power at 110\,mK. As observed, both devices exhibited an initial increase in $Q$ at low power levels, consistent with the saturation of two-level system losses. With further increases in $P_r$, the quality factor decreased, indicating the onset of quasiparticle heating and the increasing dominance of non-equilibrium quasiparticle losses induced by the readout signal. This behavior was qualitatively consistent with previous reports on superconducting resonators~\cite{thomas2020nonlinear, skyrme2023superconducting}. 

Fig.~\ref{fig:S1_measured_S21_vs_bath_temperature_at_different_readout_power_R2}(a) shows the measured transmission magnitude $|S_{21}|$ versus bath temperature for R2 at -50\,dBm readout power, where TLS losses were nearly saturated.  The resonance, initially measured at 3.4290\,GHz with $Q_r \sim$ 3400 and $Q_c \sim$21000 at 102\,mK, progressively broadened and decreased in amplitude as temperature increased to 242\,mK, before finally disappearing entirely at higher temperatures. Fig.~\ref{fig:S1_measured_S21_vs_bath_temperature_at_different_readout_power_R2}(b) shows the extracted quality factor as a function of temperature at different readout power levels, and the dashed curves correspond to fits using the model developed in Sec.\ref{sec_S2_theoretical_model}. The model provided good agreement with the experimental data across the full temperature range, capturing both the low-temperature saturation and the high-temperature exponential decay of the Q-factor.

Similar temperature sweep measurements were performed on resonator R1 fabricated on the same chip, which exhibited a resonance at 3.7895\,GHz with $Q_r \sim$2519 and $Q_c \sim$19000 at 110\,mK, as shown in Fig.~\ref{fig:S1_measured_S21_vs_bath_temperature_at_different_readout_power_R1}. The measured Q-factor versus bath temperature was again well described by the model, with $Q_{other}$ as the sole best-fit outcome. As shown in both Fig.~\ref{fig:S1_measured_S21_vs_bath_temperature_at_different_readout_power_R2} and Fig.~\ref{fig:S1_measured_S21_vs_bath_temperature_at_different_readout_power_R1}, the model demonstrated strong agreement over a readout power dynamic range of up to 40\,dB and across temperatures reaching approximately $T_c/2$. Overall, across both devices, this close agreement indicates that the modified Rothwarf-Taylor framework captures the dominant loss mechanisms governing device performance.

\section{Conclusion}\label{sec_Conclusion}
In this work, we developed a macroscopic framework for describing the quality factor of superconducting resonators as a joint function of bath temperature and microwave readout power. By extending the Rothwarf-Taylor equations to include a power-dependent quasiparticle generation term, the model provides a direct link between nonequilibrium quasiparticle dynamics and experimentally accessible quality factors. The formulation avoids an effective-temperature description and instead tracks the quasiparticle density explicitly, enabling a transparent interpretation of temperature- and power-dependent dissipation.

The model was tested against systematic bath-temperature sweeps performed at multiple readout power levels on NbN microstrip resonators with $\beta$-Ta terminations. Across a wide 40\,dB dynamic range in power and up to temperatures approaching $T_c/2$, the framework consistently captured the low-temperature plateau, the crossover regime, and the high-temperature exponential decay of the quality factor. The agreement obtained with a single free loss parameter indicates that the dominant dissipation mechanisms are adequately described within the modified Rothwarf-Taylor picture. The proposed framework is therefore valuable as a practical tool for guiding operating conditions and device design in high-Q superconducting resonators used for quantum sensing and microwave quantum information applications. Future studies should evaluate the driven temporal dynamics of superconducting resonators against the modified Rothwarf-Taylor model with the quasiparticle generation term introduced in this work.

\begin{acknowledgments}
The authors thank Dr. Thomas for proposing the concept of the theoretical framework and his guidance to Dr. Sun throughout the experimental investigations.
\end{acknowledgments}

\bibliographystyle{iopart-num}
\bibliography{reference}

\end{document}